\documentclass[aps,prb,twoside,twocolumn,floatfix,reprint]{revtex4-1}

\usepackage{amsmath,amsfonts}
\usepackage{mathptmx}
\usepackage{graphicx}

\usepackage[colorlinks,linkcolor=blue,citecolor=blue,urlcolor=blue]{hyperref}
\newcommand{\eps}{\ensuremath{\varepsilon}}
\newcommand{\del}{\ensuremath{\partial}}
\newcommand{\abs}[1]{\ensuremath{\lvert\mkern1mu{#1}\mkern1mu\rvert}}

\renewcommand{\Im}{\operatorname{Im}}
\renewcommand{\Re}{\operatorname{Re}}

\usepackage{fancyhdr}
\begin{document}

\setcounter{page}{476} %% 476--484

\title{Electromagnetic surface wave propagation in a metallic wire and the Lambert $W$ function}

\author{J. Ricardo G. Mendon\c{c}a}
\email{jrgmendonca@lptms.u-psud.fr}

\affiliation{\mbox{LPTMS, UMR 8626, CNRS, Universit\'{e} Paris-Sud, Universit\'{e} Paris-Saclay, 91405 Orsay CEDEX, France}}

\altaffiliation{\footnotesize Permanent address: Escola de Artes, Ci\^{e}ncias e Humanidades, Universidade de S\~{a}o Paulo, Rua Arlindo B\'{e}ttio 1000, 03828-000 S\~{a}o Paulo, SP, Brazil. Email:~\href{mailto:jricardo@usp.br}{jricardo@usp.br}.}

\begin{abstract}
We revisit the solution due to Sommerfeld of a problem in classical electrodynamics, namely, that of the propagation of an electromagnetic axially symmetric surface wave (a low-attenuation single TM$_{01}$ mode) in a cylindrical metallic wire, and his iterative method to solve the transcendental equation that appears in the determination of the propagation wave number from the boundary conditions. We present an elementary analysis of the convergence of Sommerfeld's iterative solution of the approximate problem and compare it with both the numerical solution of the exact transcendental equation and the solution of the approximate problem by means of the Lambert $W$ function.
\hfill \href{https://doi.org/10.1119/1.5100943}{https://doi.org/10.1119/1.5100943}
\\[12pt]
\end{abstract}

% PACS number(s):
% 41.20.-q Applied classical electromagnetism
% 41.20.Jb Electromagnetic wave propagation; radiowave propagation
% 02.30.-f Function theory, analysis

\maketitle

%% %% %% %% %% %% %% %% %% %% %% %% %% %% %% %% %% %% %% %% %% %% %% %% %% %% %%

\section{\label{sec:intro}Introduction}

The propagation of electromagnetic (EM) waves across metallic wires was a possibility of considerable interest in the second half of the XIX century, due both to the advent of the modern theory of electromagnetism and the deployment of the early electric telegraph networks.\cite{Huurdeman2003,Nahim2002} First attempts to investigate EM waves on wires were made, among others, by Hertz during 1886--1887 using spark gaps as sources of EM radiation. Some of the main questions at that time were whether EM waves propagate with a finite velocity, as predicted by Maxwell's equations, and, if so, whether they propagate at the same speed on air and other materials. Underlying these questions was the larger issue of the similarity of EM waves to light. Hertz could solve neither the theoretical nor the experimental problems involving wires, despite his many other outstanding contributions to the field.\cite{Hertz1893,Sengupta2003}

In 1899, Sommerfeld tackled the problem of the propagation of a circular symmetric EM wave on a thin conducting wire using the full set of Maxwell's equations, demonstrating that such a cylindrical geometry could support what has become known as a surface wave.\cite{Sommerfeld1899} He showed, by means of an example (see Sec.~\ref{sec:wire}), that the attenuation of the surface EM waves at high frequencies was too large to make them useful for telecommunications. In the ensuing years, Sommerfeld's work was extended to include noncircular symmetric modes and to treat coated wires and dielectric cylinders,\cite{Harms1907,Hondros1909,Debye1910,Schriever1920} untill by the mid-1930s the general interest shifted to the study of EM cavities (metal boxes) and waveguides (hollow metal tubes) as more practical and efficient devices to store and transmit EM energy across space.\cite{Southworth1936,Southworth1937,Collin1999,Carson1936,Barrow1937,Packard1984}

% TM (TE) mode is a wave in which the magnetic (electric) field vector $\mathbf{H}$ ($\mathbf{E}$) is perpendicular to the direction of wave propagation (equivalently, the component of the field vanishes along the waveguide)

% There are also solutions describing periodic waves around the cylinder. These solutions have been investigated by D.~Hondros \cite{}. However, most of the field energy of these nonsymmetrical modes is inside the conductor and they are heavily attenuated, making them of little practical interest.

% In addition, higher-order modes, with angular dependence, may exist on a conducting wire. These modes, however, attenuate so rapidly with distance that they are of no importance except as part of the near-zone field at the region of the exciting source.

In $1950$, a German physicist by the name of Georg J. E. Goubau, brought to USA from Germany after World War II as part of Operation Paperclip, resumed the investigation of the ``Sommerfeld waves'' with regard to their practical application to transmission lines.\cite{Goubau1950} Goubau was able to show, both theoretically and experimentally, that EM surface waves can be generated on wires by means of coaxial flared horn-shaped launchers, making the otherwise theoretical waves practical. His main result, though, was the demonstration that a sheathed or subwavelength corrugated wire would perform better than a naked wire in the transmission of low frequency ($<1$~THz) EM energy along the line. This is not because of substantially less ohmic or dielectric losses, but because the modified surface of the wire has a much lower effective conductivity than the metal alone, reducing the extension of the radial electric field outside the wire (field confinement), which for low frequencies and large wire diameters can be large. For example, a wire of radius $a=10$~mm carrying a surface wave of $1$~GHz propagates $75\%$ of its power into an area of radius $\sim 1.5$~m around the wire,\cite{Goubau1950} not only diverting energy but also potentially coupling with the environment and compromising the entire engineering solution by requiring too much clearance around the transmission line. Surface EM waves on wires have since been known in radio engineering as Sommerfeld-Goubau waves. \cite{BarlowCullen1953a,BarlowCullen1953b,Roberts1953,Barlow1958} Such waves have repeatedly been considered as a cheap alternative to long-distance, beyond-the-horizon transportation of EM energy---for example, for last mile simplex or half-duplex data communication over existing overhead electric power lines---in conditions under which microwave and coaxial links become inefficient or costly.\cite{FriedFern2001,WangMitt2004,JeonZhangGris2005,Akalin2006}

% Subwavelength corrugations on metal waveguides are known as ``spoof plasmons'' in the modern literature.\cite{Maier2007}

In this paper we revisit Sommerfeld's solution of the problem of an EM axially symmetric surface wave (to wit, a pure transverse magnetic, TM$_{01}$ mode) propagating in a cylindrical metallic wire. Of particular interest to us is the solution of the boundary condition for the EM fields that lead to a transcendental equation that Sommerfeld solved approximately by means of a practical and effective device. We explore the iterative procedure devised by Sommerfeld to solve the transcendental equation and briefly investigate its convergence. As we shall see, the solution of the approximate physical system can immediately be given in terms of the (complex-valued) Lambert $W$ function, a most interesting function that gained a lot of attention in the last few decades. Both solutions (Sommerfeld's and the one in terms of the Lambert $W$ function) are compared with the exact numerical solution of the boundary condition in a concrete case. Some remarks on the physical significance of the solutions are made.

%% %% %% %% %% %% %% %% %% %% %% %% %% %% %% %% %% %% %% %% %% %% %% %% %% %% %%

\vspace{-3pt}
\section{\label{sec:wire}Circular symmetric EM wave \\ in a metallic wire}

Comprehensive treatments of EM waves particularly suited to our discussion can be found in Refs.~\onlinecite[Chap.~IX]{Stratton1941} and \onlinecite[$\S{20}$--$\S{25}$]{Sommerfeld1952}, while for definitions and properties of special functions we refer the reader to Refs.~\onlinecite{SommerfeldVI,Arfken7e,NIST2018}. Except for one mention to electron-volts in Sec.~\ref{sec:exact}, we use SI units throughout.

Maxwell's equations in a conducting medium without free charges lead to the following wave equation for the electric field $\mathbf{E}=\mathbf{E}(\mathbf{r},t)$,
\begin{equation}
\label{eq:maxwave}
\nabla^{2}\mathbf{E}
-{\sigma\mu}\frac{\del\mathbf{E}}{\del t}
-{\eps\mu}\frac{\del^{2}\mathbf{E}}{\del t^{2}}=0,
\end{equation}
with an identical equation for the magnetic field intensity $\mathbf{H}=\mathbf{H}(\mathbf{r},t)$, where $\eps$ and $\mu$ are, respectively, the time-independent electric permittivity and magnetic permeability of the material and $\sigma$ its electrical conductivity. Outside of the conductor $\sigma = 0$, and in vacuum (and to a very good approximation also dry air) $\eps=$ $\eps_{0}=$ $1/\mu_{0}c^{2}$~F\,m$^{-1}$ $\simeq 8.854 \times 10^{-12}$~F\,m$^{-1}$ and $\mu=$ $\mu_{0}=$ $4\pi \times 10^{-7}$~H\,m$^{-1}$, with $c = 1/\sqrt{\eps_{0}\mu_{0}} = 299\,792\,458$~m\,s$^{-1}$ the exact speed of light.

\begin{figure}[t]
\centering
\includegraphics[viewport=140 160 720 480, scale=0.40, clip]{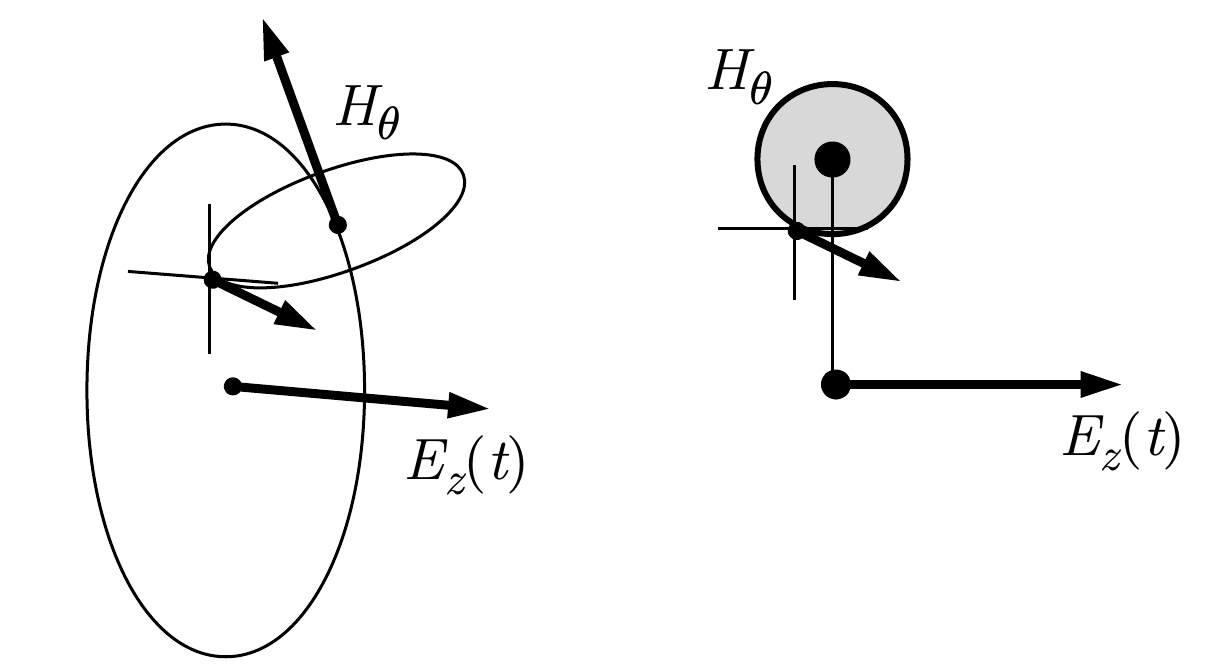}
\caption{Schematic geometry of the fields inside the wire, in perspective (left panel) and in the $z$-$r$ plane (right panel), where $H_{\theta}$ is depicted coming out of the page.}
\label{fig:geometry}
\end{figure}

Let us consider a long and thin metallic wire of cylindrical symmetry and constant circular cross section of radius $a$ embedded in a dielectric medium, and assume that there is only a harmonic oscillating electric field along the direction of propagation $z$, i.\,e., a pure transverse magnetic (TM) mode. This kind of EM wave can be excited on one end of the wire provided that its source is infinitely far removed, or by some device specifically designed for the task, as the coaxial horn-shaped launchers devised by Goubau in the 1950s (essentially mode transducers transforming the TEM wave of a coaxial cable to the TM$_{01}$ wave of the surface waveguide).\cite{Goubau1950,RTVNews1955} Following the usual practice, we introduce the complex wave number $h$ (to be determined from the boundary conditions) to express both the propagation and the attenuation of the EM wave across the wire in the $z$-direction through the time-dependent phase factor $\phi(z,t)=\exp[i(hz-{\omega}t)]$ with a real angular frequency $\omega$, such that
\begin{equation}
\label{eq:phasor}
\mathbf{E}(\mathbf{r},t) = \mathbf{E}^{0}(r,\theta)\phi(z,t) = \mathbf{E}^{0}(r,\theta)\exp[i(hz-{\omega}t)]
\end{equation}
and similarly for $\mathbf{H}(\mathbf{r},t)$. The imaginary part of $h$ must be positive if the wave is to attenuate; we shall also take the real part of $h$ positive, corresponding to the phase of the wave propagating in the positive $z$-direction. Because of the cylindrical symmetry of the problem, the time-independent field vector $\mathbf{E}^{0}(r,\theta)$ depends only on the $r$-coordinate, $\mathbf{E}^{0}(r,\theta)=\mathbf{E}^{0}(r)$, and similarly for $\mathbf{H}^{0}(\mathbf{r})$. Within these settings, Maxwell's equations
\begin{subequations}
\label{eq:sEmH}
\begin{equation}
\nabla\times\mathbf{H} = \sigma\mathbf{E}+\frac{\del(\eps\mathbf{E})}{\del t} \quad (\text{Amp\`{e}re's circuital law})
\end{equation}
and
\begin{equation}
\nabla\times\mathbf{E} = -\frac{\del(\mu\mathbf{H})}{\del t} \quad (\text{Faraday's law of induction})
\end{equation}
\end{subequations}
imply that an electric field $E_{z}$ varying along the $z$-direction induces (by Amp\`{e}re's law) a circulating magnetic field $H_{\theta}$ that, in turn, induces (by Faraday's law) an electric field with components in the $r$ and $z$ directions---and the three fields $E_{z}$, $H_{\theta}$, and $E_{r}$ suffice to describe the physics of the problem. The other components $E_{\theta}$, $H_{z}$, and $H_{r}$ vanish together. The geometry of these fields is depicted schematically in Fig.~\ref{fig:geometry}.

The physics of the problem indicates that everything must depend on $E_{z}$, the sole forcing field. Indeed, perusal of Eqs.~(\ref{eq:sEmH}) shows that $E_{r}^{0}$ and $H_{\theta}^{0}$ can be obtained from $E_{z}^{0}$ as
\begin{subequations}
\label{eq:ErHt}
\begin{align}
\label{eq:Er}
E_{r}^{0}(\hat{\rho}) &= \frac{ih}{\sqrt{\hat{k}^{2}-h^{2}}}
\frac{\del E_{z}^{0}(\hat{\rho})}{\del\hat{\rho}}, \\
\label{eq:Htheta}
\sqrt{\frac{\mu}{\hat{\eps}}} H_{\theta}^{0}(\hat{\rho}) &=
\frac{i\hat{k}}{\sqrt{\hat{k}^{2}-h^{2}}}
\frac{\del E_{z}^{0}(\hat{\rho})}{\del\hat{\rho}},
\end{align}
\end{subequations}
where we have introduced the complex quantities
\begin{subequations}
\label{eq:complex}
\begin{align}
\label{eq:hatk}
\hat{k}^{2} &= \eps\mu\omega^{2}+i\sigma\mu\omega, \\
\label{eq:hatrho}
\hat{\rho} &= \sqrt{\hat{k}^{2}-h^{2}}\,r, \\
\label{eq:hateps}
\hat{\eps} &= \frac{\hat{k}^{2}}{\mu\omega^{2}} = \eps+i\,\frac{\sigma}{\omega}.
\end{align}
\end{subequations}
The hats indicate the quantities inside the wire ($0<r<a$), where $\sigma \ne 0$; outside of the wire ($r>a$) they become real and lose their hat. For definiteness, we always take the imaginary part of $\hat{\rho}$ to be positive, which unambiguously defines the sign of the square root in Eq.~(\ref{eq:hatrho}). The introduction of the complex dielectric constant $\hat{\eps}=\hat{\eps}(\omega)$ allows us to unify the treatment of the problem for dielectric media, conducting or not. In particular, it allows us to write the complex dispersion relation
\begin{equation}
\label{eq:dispersion}
\hat{k}(\omega)=\sqrt{\hat{\eps}\mu}\,\omega
\end{equation}
everywhere. If the medium is nonconducting, $\hat{\eps}=\eps$ and the usual dispersion relation $k(\omega)=\sqrt{\eps\mu}\,\omega=(n/c)\omega$ recovers, with $n \geq 1$ the refractive index of the medium. Definition (\ref{eq:hateps}) was employed in the simplification of Eq.~(\ref{eq:Htheta}).

We must find $E_{z}^{0}(\hat{\rho})$ to complete the description of the system. It follows from Eq.~(\ref{eq:maxwave}) that $E_{z}^{0}$ obeys
\begin{equation}
\label{eq:EzJ0}
\frac{d^{2}E_{z}^{0}}{d\hat{\rho}^{2}}+
\frac{1}{\hat{\rho}}\frac{dE_{z}^{0}}{d\hat{\rho}}+E_{z}^{0}=0.
\end{equation}
Equation (\ref{eq:EzJ0}) is a particular case ($\nu=0$) of Bessel's differential equation, for which the last term on the left-hand side in the general case would read $(1-\nu^{2}/\hat{\rho}^{2})E_{z}^{0}$, with $\nu$ a complex number. The physical solutions of Eq.~(\ref{eq:EzJ0}) are given~by
\begin{subequations}
\label{eq:Ez}
\begin{align}
\label{eq:Ezint}
E_{z}^{0}(\hat{\rho}) &=
A\frac{\sqrt{\hat{k}^{2}-h^{2}}}{ih}\,J_{0}(\hat{\rho}), \quad 0 < r < a, \\
\label{eq:Ezext}
E_{z}^{0}(\rho) &=
B\frac{\sqrt{k^{2}-h^{2}}}{ih}\,H_{0}^{(1)}(\rho), \quad a < r < \infty,
\end{align}
\end{subequations}
where $A$ and $B$ are arbitrary real constants and $J_{0}(\hat{\rho})$ and $H_{0}^{(1)}({\rho})$ are, respectively, Bessel and Hankel functions of the first kind. The other field components follow from Eq.~(\ref{eq:ErHt}).

Solution (\ref{eq:Ezint}) contains only the regular Bessel function, because the other linearly independent solution---the Neumann function---is singular at the origin. Recently, it has been noticed that solution (\ref{eq:Ezint}) (including the phase factor) can actually be split into two counter-propagating conical waves each containing the Neumann function in its description.\cite{Correa2017} The net result, however, coincides with Eq.~(\ref{eq:Ezint}). Solution (\ref{eq:Ezext}), however, could in principle read as a linear combination of the Hankel functions of the first and second kinds,
\begin{equation}
\label{eq:H1H2}
E_{z}^{0}(\rho) = B_{1}H_{0}^{(1)}(\rho) + B_{2}H_{0}^{(2)}(\rho), \quad a < r < \infty.
\end{equation}
We dismiss this possibility on the basis of two arguments. First, note that the leading asymptotic form ($\abs{\rho} \gg 1$, $\abs{\nu} \ll \abs{\rho}$) of the Hankel functions is
\begin{equation}
\label{eq:hankel}
H_{\nu}^{(1,2)}(\rho) \sim \sqrt{\frac{2}{\pi\rho}}\exp\big[\pm i(\rho-\tfrac{1}{2}\nu\pi-\tfrac{1}{4}\pi)\big],
\end{equation}
where the $+$ ($-$) sign in the exponent belongs to $H_{\nu}^{(1)}$ ($H_{\nu}^{(2)}$). If $\rho=\sqrt{k^{2}-h^{2}}$ is complex, then, as per our sign convention (see discussion following Eq.~(\ref{eq:complex})), we must discard $H_{0}^{(2)}(\rho)$ because it diverges as $\Im{\rho} \to \infty$. If, otherwise, $\rho$ is real, an energy condition rules $H_{0}^{(2)}(\rho)$ out from the solution as follows. For an infinite wire sustaining a longitudinal EM wave that draws its energy from a source at $z=-\infty$, the energy flux across a cylindrical surface coaxial with the wire must vanish. This energy flux per unit length is given by $U_{r}(r,t)=2\pi rS_{r}(r,t)$, where $S_{r}$ is the radial component of the Poynting vector $\mathbf{S}=\mathbf{E} \times \mathbf{H}$, namely,
\begin{equation}
S_{r}(r,t)=-\Re\big\{E_{z}^{0}(r)\phi(z,t)\big\} \Re\big\{H_{\theta}^{0}(r)\phi(z,t)\big\}.
\end{equation}
Taking Eq.~(\ref{eq:hankel}) into account and doing the algebra we obtain
\begin{widetext}
\begin{equation}
U_{r}(r,t) = 4\frac{r}{\rho} \sqrt{\frac{\eps_{0}}{\mu_{0}}} \frac{k}{\sqrt{k^{2}-h^{2}}} \Big[B_{1}^{2}\cos^{2}(\rho-\tfrac{1}{4}\pi+\text{phase})-B_{2}^{2}\cos^{2}(\rho-\tfrac{1}{4}\pi-\text{phase})\Big].
\end{equation}
\end{widetext}
It is always healthy to check once in a while (particularly in lengthy calculations or derivations with many intermediate steps) if the physical dimensions of the expressions obtained are correct. For example, at this point it is a good idea to verify that $U_{r}(r,t)$ indeed has dimensions of energy flux per unit length, [$U_{r}$] $=$ MLT$^{-3}$ $=$ $($ML$^2$T$^{-2}/$T$)/$L (in SI units, J\,s$^{-1}$\,m$^{-1}$). Since $r/\rho$ remains finite (in fact, constant) at arbitrarily large $r$ and the phase $hz-\omega t$ can assume any value we conclude that, for real $\rho$, the condition $U_{r}(r \gg a,t)=0$ leads to $B_{1}=0=B_{2}$. Equations (\ref{eq:Ez}) are then justified.

Finally, fields $E_{z}^{0}$ and $H_{\theta}^{0}$ must be continuous at the surface ($r=a$) of the wire. This boundary condition leads to the transcendental equation
\begin{equation}
\label{eq:trans}
\begin{split}
\sqrt{\frac{\mu_{0}}{\eps_{0}}} \,\frac{\sqrt{k^{2}-h^{2}}}{k} \,
&\frac{H_{0}^{(1)}\big(a\sqrt{k^{2}-h^{2}}\big)}{H_{0}^{(1)\,\prime}\big(a\sqrt{k^{2}-h^{2}}\big)} = \\
&=\sqrt{\frac{\mu}{\hat{\eps}}} \,\frac{\sqrt{\hat{k}^{2}-h^{2}}}{\hat{k}} \,
\frac{J_{0}\big(a\sqrt{\hat{k}^{2}-h^{2}}\big)}{J_{0}^{\,\prime}\big(a\sqrt{\hat{k}^{2}-h^{2}}\big)},
\end{split}
\end{equation}
where the primes indicate differentiation with respect to the given argument. Equation (\ref{eq:trans}) determines the wave number $h$, from which most of the electromagnetic properties of the system can be derived. In what follows we solve Eq.~(\ref{eq:trans}) both numerically and by an approximate method due to Sommerfeld and compare the results.

%% %% %% %% %% %% %% %% %% %% %% %% %% %% %% %% %% %% %% %% %% %% %% %% %% %% %%

\section{\label{sec:exact}Exact numerical solutions of the boundary condition}

Before embarking on the solution of Eq.~(\ref{eq:trans}), let us look at the magnitude of the physical quantities involved in a typical case.

Consider a microwave propagating in a hard-drawn copper wire of radius $a=1$~mm (approximately wire no.~$12$ in the American Wire Gauge system, for which $2a=2.053$~mm). The following parameters are taken from Refs.~\onlinecite{AshcroftMermin1976,Wikipedia2018}. The dc electrical conductivity of copper at $293$~K ($\sim 20$~$^{\circ}$C) is $5.96 \times 10^{7}$~$\Omega^{-1}$\,m$^{-1}$, while for $f=1$~GHz we have $ka \simeq 0.021$, where $k=\omega/c = 2\pi f/c = 2\pi/\lambda$ is the free-space EM wave number and $\lambda=c/f \simeq 30$~cm its wavelength. Note that for most metals the electrical conductivity is essentially static up to about the infrared region $f \sim 10$~THz or, more meaningfully, photon energies about $E = hf \sim 40$~meV (where Planck's constant $h \simeq 4.136 \times 10^{-15}$~eV\,s);\cite{planck} microwaves ($f \sim 1$--$100$~GHz, $\lambda \sim 300$--$3$~mm, $E \sim 0.004$--$0.4$~meV) are far below this energy scale. As we have already mentioned, for copper $\mu \simeq \mu_{0}$ (the more precise current value is $\mu_{\mathrm{Cu}}=0.999\,994\,\mu_{0}$), but the electric permittivity is a complex-valued function of the frequency, $\hat{\eps}=\hat{\eps}(\omega)$. For most metals, however, the conduction current $\mathbf{J}=\sigma\mathbf{E}$ is much larger than the displacement current $\mathbf{J}_{D}=\eps\del_{t}\mathbf{E} = -i\omega\eps\mathbf{E}$, since $\eps/\sigma \ll 1$. In fact, for most metals $\eps/\sigma \sim 10^{-18}$, and as long as $f \lesssim 10^{16}$~Hz the displacement current can be ignored. This corresponds to the low-frequency limit of an ac Drude model for copper. We thus obtain from Eq.~(\ref{eq:hatk}) that $\hat{k}^{2} \approx i\sigma\mu\omega$, or
\begin{equation}
\label{eq:kappa}
\hat{k} \approx \frac{1+i}{\sqrt{2}}\sqrt{\sigma\mu\omega} = (1+i)\kappa(\omega).
\end{equation}
The real quantity $\delta(\omega)=1/\kappa(\omega)$ measures the depth in the material at which the current density decreases to a fraction $e^{-1} \simeq 37\%$ of its value at the surface---the so-called skin depth. Note that $\delta(\omega) \sim 1/\sqrt{\omega}$ can become a really small quantity at high frequencies, sometimes just a few nanometers, with the current mostly confined to the periphery of the conductor. If we plug in the figures of our example we obtain $\kappa \simeq 4.851 \times 10^{5}$~m$^{-1}$, that is, $\delta \simeq 2~\mu$m.

To find the wave number $h$, we solve Eq.~(\ref{eq:trans}) numerically with the help of the software package Mathematica\cite{Wolfram2018} using the figures given above for the physical quantities, as well as approximation (\ref{eq:kappa}) for $\hat{k}$. Graphing the two sides of Eq.~(\ref{eq:trans}) is unilluminating, so we just quote the solution,
\begin{equation}
\label{eq:exacth}
h = 20.960+7.516 \times 10^{-4}\,i.
\end{equation}
Two conclusions of physical significance can be immediately drawn from this number. First, that the wave is damped, since the phase factor $\exp(ihz)$ contains an attenuation factor given by $\exp(-7.516 \times 10^{-4}z)$, and the amplitude of the wave decays by a factor $e^{-1}$ every $\sim 1330$~m. This attenuation can be considered small, and such a naked copper wire no.~12 at $1$~GHz indeed qualifies as a viable candidate for last mile data carrier! The second conclusion is that the wave travels along the wire at phase speed
\begin{equation}
\frac{\omega}{h} = \frac{\omega}{k} \cdot \frac{k}{h} = c[1-(7.438+3.586\,i) \times 10^{-5}],
\end{equation}
i.\,e., at about $0.999\,926\,c$, just a tad below the speed of light. The ``surface wave'' is thus able to travel without much damping for a reasonable distance---but not telegraphic distances, much to the chagrin of the telecommunications engineers at the turn of the XX century.

%% %% %% %% %% %% %% %% %% %% %% %% %% %% %% %% %% %% %% %% %% %% %% %% %% %% %%

\section{\label{sec:approx}Sommerfeld's approximate solution}

Sommerfeld approached the solution of Eq.~(\ref{eq:trans}) by making well informed assumptions about the physical nature of the system at hand and then approximating the functions appearing there accordingly. He was then led to a much simpler transcendental equation that he solved by means of a practical and effective device. In what follows we review the rationale leading to the approximate equation for the boundary condition and Sommerfeld's iterative solution to the problem.

\subsection{\label{sec:physical}The physical approximations}

For a metallic wire, $\sigma$ is large and $\hat{k}$ and $\hat{\rho}$ are large complex numbers. Given that $J_{0}^{\,\prime}(\hat{\rho}) = -J_{1}(\hat{\rho})$ and that for large $\hat{\rho}$ in the complex upper half-plane (remember that we are using $\Im\hat{\rho} \geq 0$ by convention) it holds that
\begin{equation}
\label{eq:bessel}
J_{\nu}(\hat{\rho}) \sim \sqrt{\frac{2}{\pi\hat{\rho}}}
\cos(\hat{\rho}-\tfrac{1}{2}\nu\pi-\tfrac{1}{4}\pi),
\end{equation}
the right-hand side of Eq.~(\ref{eq:trans}) can be simplified by taking $\sqrt{\hat{k}^{2}-h^{2}}/\hat{k} \approx 1$ and $J_{0}(\hat{\rho})/J_{0}^{\,\prime}(\hat{\rho}) \approx -e^{-i\frac{\pi}{2}} = i$ (we always take the principal branch), such that
\begin{equation}
\label{eq:rhs}
\sqrt{\frac{\mu}{\hat{\eps}}} \,
\frac{\sqrt{\hat{k}^{2}-h^{2}}}{\hat{k}} \,
\frac{J_{0}\big(a\sqrt{\hat{k}^{2}-h^{2}}\big)}{J_{0}^{\,\prime}\big(a\sqrt{\hat{k}^{2}-h^{2}}\big)} \simeq i\sqrt{\frac{\mu}{\hat{\eps}}}.
\end{equation}

In metals widely used for electric wiring like copper, silver or aluminium, $\abs{\mu} \simeq \mu_{0}$ to a very good approximation but $\abs{\hat{\eps}} \gg \eps_{0}$, and we see from Eq.~(\ref{eq:rhs}) that the left-hand side of Eq.~(\ref{eq:trans}) must also be a small number, i.\,e., $\abs{\!\sqrt{k^{2}-h^{2}}}$ must be small. The expansion of the Hankel functions about the origin gives
\begin{subequations}
\label{eq:Hankel}
\begin{align}
H_{0}^{(1)}(\rho) &\approx \frac{2i}{\pi}\ln\Big(\frac{e^{\gamma}\rho}{2i}\Big), \\
H_{\nu}^{(1)}(\rho) &\approx \frac{\Gamma(\nu)}{i\pi}\Big(\frac{2}{\rho}\Big)^{\nu}, \quad \nu > 0,
\end{align}
\end{subequations}
where $\gamma=0.577\cdots$ is the Euler-Mascheroni constant and $\Gamma(\nu)$ is the usual gamma function. Since $H_{0}^{\,\prime}(\rho) = -H_{1}(\rho)$, the left-hand side of Eq.~(\ref{eq:trans}) becomes
\begin{equation}
\label{eq:lhs}
\begin{split}
\sqrt{\frac{\mu_{0}}{\eps_{0}}} \, &\frac{\sqrt{k^{2}-h^{2}}}{k} \,
\frac{H_{0}\big(a\sqrt{k^{2}-h^{2}}\big)}{H_{0}'\big(a\sqrt{k^{2}-h^{2}}\big)} \simeq \\
&\simeq \sqrt{\frac{\mu_{0}}{\eps_{0}}} \,\frac{a(k^{2}-h^{2})}{k} \,
\ln\Big(\frac{e^{\gamma}a\sqrt{k^{2}-h^{2}}}{2i}\Big).
\end{split}
\end{equation}

Now we put everything together. Defining the variables
\begin{equation}
\label{eq:uv}
u = \Big(\frac{e^{\gamma}a\sqrt{k^{2}-h^{2}}}{2i}\Big)^{2}
\quad \mathrm{and} \quad
v = \frac{e^{2\gamma}}{2i}\sqrt{\frac{\eps_{0}\mu}{\hat{\eps}\mu_{0}}}\,ka,
\end{equation}
Eq.~(\ref{eq:trans}) for the wave number $h$ of a relatively low frequency pure TM mode propagating in a cylindrical metallic wire becomes, within the aforementioned approximations,
\begin{equation}
\label{eq:ulnuv}
u \ln{u} = v.
\end{equation}
This equation corresponds to equations ($22$)--($22'$) in Sommerfeld's original paper,\cite{Sommerfeld1899} equations ($9$)--($10$) in Goubau's treatment of the problem,\cite{Goubau1950} equations (IX.$23$--$24$) in the textbook by Stratton,\cite{Stratton1941} and equation ($22.15$) in in the textbook by Sommerfeld.\cite{Sommerfeld1952} Note that while on the left-hand side of Eq.~(\ref{eq:ulnuv}) we have the geometric factors of the problem ($a$, $k$, and $h$), the right-hand side concentrates the physical factors ($\hat{\eps}/\eps_{0}$ and $\mu/\mu_{0}$), providing a clear illustration of the role played by boundary conditions in electrodynamics.

\subsection{\label{sec:iterative}The iterative solution}

Sommerfeld solved Eq.~(\ref{eq:ulnuv}) by ``a method of successive approximations [that] proved to be very suitable.''\cite{Sommerfeld1898} Several references repeat his solution, sometimes second-hand from Stratton,\cite{Stratton1941} while Goubau offers an interesting approximate graphical solution of his own.\cite{Goubau1950} In what follows we present Sommerfeld's method according to more modern standards.

% Bei Gelegenheit einer Aufgabe aus der Elektrodynamik wurde ich auf eine transcendente Gleichung gef\"{u}hrt, f\"{u}r deren numerische Aufl\"{o}sung sich ein Verfahren successiver Approximationen sehr geeignet erwies.
% By the occasion of an exercise in electrodynamics, I was led to a transcendental equation for whose numerical solution a method of successive approximations proved to be very suitable.

\begin{figure}[b]
\centering
\includegraphics[viewport=20 60 540 430, scale=0.40, clip]{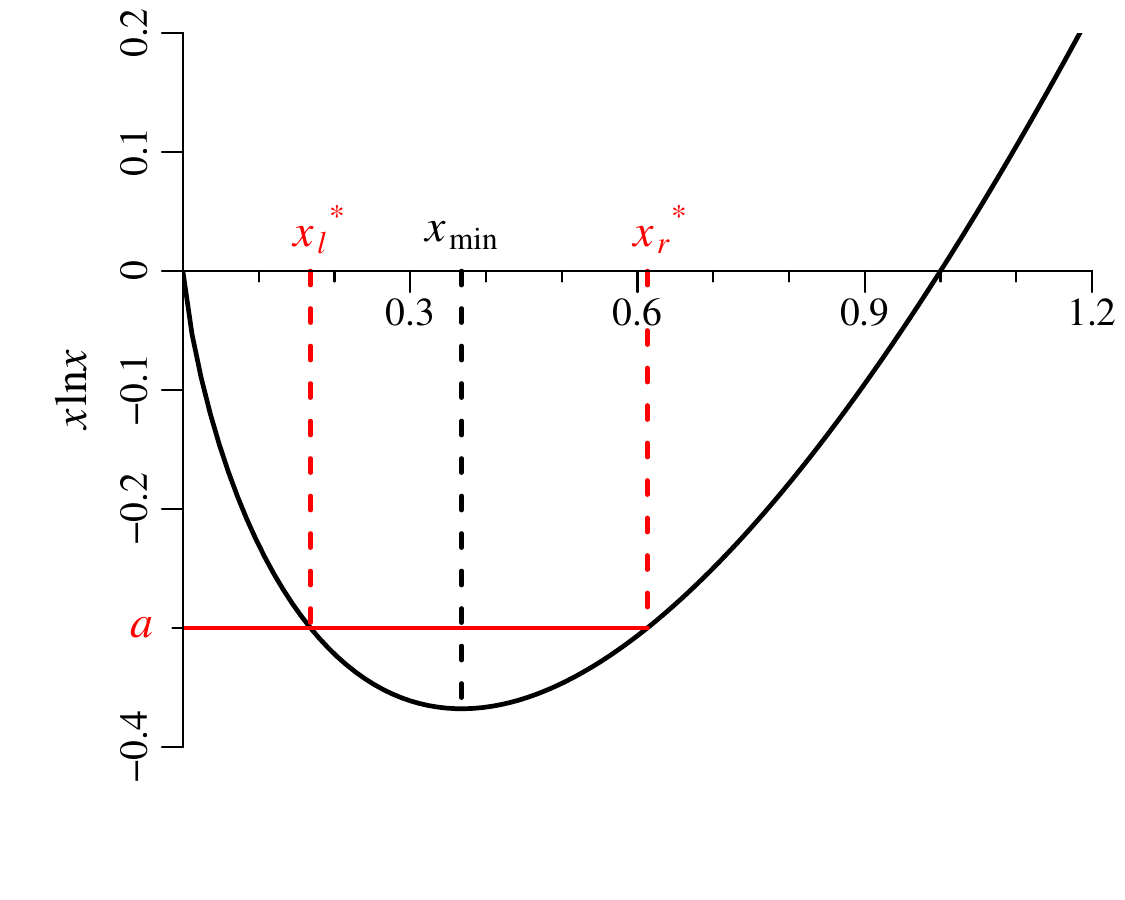}
\caption{Location of the solutions of $x\ln{x}=a$ for $-e^{-1} < a < 0$.}
\label{fig:xlnxa}
\end{figure}

Let $f(x)=x\ln{x}$ be a real function of a real argument $x \geq 0$. Routine inspection reveals that $f(x)$ has an absolute minimum at $x_{\text{min}}=e^{-1} \simeq 0.367\,879$, at which $f(x_{\text{min}})=-e^{-1}$, and since $\lim_{x \to 0}f(x)=0$ and $f(1)=0$, it is decreasing for $0 \leq x < e^{-1}$ and increasing for $x > e^{-1}$. We thus conclude that the equation
\begin{equation}
\label{eq:xlnxa}
x\ln{x}=a
\end{equation}
does not have real solutions for $a < -e^{-1}$, has two solutions for $-e^{-1} \leq a < 0$, one in the interval $(0,e^{-1})$ and the other in the interval $(e^{-1},1)$ (the two coinciding when $a=-e^{-1}$), and has a single solution for $a \geq 0$. These facts are summarized in Fig.~\ref{fig:xlnxa}.

\subsubsection{\label{sec:smaller}The smaller solution}

Based on the previous analysis, Sommerfeld devised two iterative procedures to find the solutions of Eq.~(\ref{eq:xlnxa}) depending on the location of the solution. Let us first consider the case $-e^{-1} < a < 0$. The smaller solution of Eq.~(\ref{eq:xlnxa}), say, $x_{l}^{*}$, is located to the left of $x_{\text{min}}=e^{-1}$. Sommerfeld then observes that if we write Eq.~(\ref{eq:xlnxa}) as $x(-\ln{x})=-a$ and take logarithms we obtain
\begin{equation}
\label{eq:xloga}
\ln{x} = \ln(-a)-\ln(-\ln{x}),
\end{equation}
and since $\ln{x}$ varies slowly in comparison with $x$ (an error of $\delta$ in the value of $x$ gives an error of roughly $\delta/x$ in the value of $\ln{x}$), we can ignore the second term on the right-hand side of Eq.~(\ref{eq:xloga}) and take $x_{1} \simeq -a$ as a first approximation to $x_{l}^{*}$. Once we have this first approximation to $x_{l}^{*}$, we can substitute it on the right-hand side of Eq.~(\ref{eq:xloga}) to obtain a new approximation, say, $x_{2}$, hopefully better than the former, given by
\begin{equation}
\label{eq:x2}
\ln{x_{2}} = \ln(-a)-\ln(-\ln{x_{1}}),~\text{ or }~x_{2}=\frac{a}{\ln{x_{1}}}.
\end{equation}
Note that since $a<0$ and $\ln(-a) < -1$ we have $0< x_{2} < x_{1}$. Plugging $x_{2}$ in Eq.~(\ref{eq:xloga}) leads to $x_{3}=a/\ln{x_{2}}$ with $0< x_{3} < x_{2}$. The iterative procedure is clear:
\begin{equation}
\label{eq:first}
x_{1}=-a, \quad x_{n+1} = \frac{a}{\ln{x_{n}}},~~n \geq 1.
\end{equation}

% The method of solving a recurrence relation by starting with some estimate and plugging it into the recurrence to derive improved estimates is generally known as bootstrapping---it is like the solution is pulling itself up by its bootstraps!

Stratton states that ``the convergence [of Eqs.~(\ref{eq:first})] can be demonstrated for complex as well as real values of $a$'' (Ref.~\onlinecite[p.~529]{Stratton1941}) but neither offers a proof nor refers to one. He was certainly quoting from Ref.~\onlinecite{Sommerfeld1899}. The mathematical work of Sommerfeld\cite{Sommerfeld1898} on the iterative solution of Eq.~(\ref{eq:ulnuv}) or (\ref{eq:xlnxa}) is rarely mentioned. Sommerfeld proved that $\lim x_{n} = x_{l}^{*}$ from Eqs.~(\ref{eq:first}) based on the ratio test and the fact that $0 < x_{n+1} < x_{n}$ to conclude that $(x_{n})$ forms a Cauchy sequence---for every given real $\eps > 0$ one can find an integer $N_{\eps}$ such that $\abs{x_{n}-x_{m}} < \eps$ for all $m,\, n \geq N_{\eps}$. We shall adopt a more general approach.

From elementary mathematical analysis,\cite{Donsig2010,Shlomo2013} we know that the ``discrete dynamical system'' (\ref{eq:first}) has a fixed point $x^{*}=T(x^{*})$ if the map $T(x)=a/\ln{x}$ is a contraction, i.\,e., if it satisfies the so-called Lipschitz condition
\begin{equation}
\label{eq:contr}
\|T(y)-T(x)\| \leq c\|y-x\|
\end{equation}
with $c<1$ for any two points $x,\, y$ in the domain of $T$. The constant $c$ is called the Lipschitz constant of the map. The nice thing about contractions is that a powerful theorem guarantees, under certain conditions, that they have a unique fixed point: the Banach contraction principle. Its statement reads:
\begin{quote}
A contraction $T \colon X \to X$ on a complete metric space $X$ has a unique fixed point $x^{*}$.
\end{quote}
Remember that a complete metric space is a metric space in which every Cauchy sequence is convergent. An easy corollary of the Banach contraction principle is that $\lim_{n \to \infty}T^{(n)}(x)=x^{*}$ for all $x \in X$, where $T^{(n)}(x)$ denotes the $n$-fold composition $(T \circ \cdots \circ T)(x)$.\cite{Hille1972}

We now revisit Eqs.~(\ref{eq:first}) in the light of the above knowledge. First, note that the interval $X=[0,e^{-1}]$ is a complete metric space (in the usual Euclidean metric $\|y-x\|=\sqrt{(y-x)^{2}}=\abs{y-x}$) because it is a closed subset of $\mathbb{R}$, which is complete. To check whether $T(x)=a/\ln{x}$ is a contraction on $X$ we take
\begin{equation}
\label{eq:lnxlny}
\abs{T(y)-T(x)} = \Big|\frac{a}{\ln{y}}-\frac{a}{\ln{x}}\Big| \leq \abs{a} \max_{x,\, y\, \in\, X}\Big|\frac{1}{\ln{y}}-\frac{1}{\ln{x}}\Big|.
\end{equation}
The largest possible value of the right-hand side of Eq.~(\ref{eq:lnxlny}) occurs when one of its terms becomes arbitrarily small (i.\,e., when $x$ or $y \to 0$) and the other becomes $1/\ln(e^{-1})=-1$, such that $\abs{T(y)-T(x)} \leq \abs{a}$ in $X$. Putting all together we get
\begin{equation}
\label{eq:bound}
\abs{T(y)-T(x)} \leq \abs{a} \leq c\max_{x,\, y\, \in\, X}\abs{y-x} = ce^{-1}.
\end{equation}
It thus suffices to have $\abs{a} < e^{-1}$ to make $T$ a contraction. Since $-e^{-1} < a < 0$, we conclude that $T(x)$ is indeed a contraction on $X=[0,e^{-1}]$ and that its unique fixed point $x^{*}=T(x^{*})$ is, by construction, the solution of Eq.~(\ref{eq:xlnxa}) on~$X$.

\subsubsection{\label{sec:larger}The larger solution}

If we try to find the second solution of Eq.~(\ref{eq:xlnxa}) in the interval $e^{-1} < x <1$ with $-e^{-1} < a < 0$ using the recursion relation in Eqs.~(\ref{eq:first}), it may or may not converge and, if it converges, it will do so to the smaller solution. From Eq.~(\ref{eq:xlnxa}), however, we see that there are two options to draw an iterative procedure of it: either $x_{n+1}\ln{x_{n}}=a$, corresponding to Eqs.~(\ref{eq:first}), or $x_{n}\ln{x_{n+1}}=a$. Sommerfeld realized that the second option provides a scheme to find the second solution of Eq.~(\ref{eq:xlnxa}) in the interval $e^{-1} < x <1$ when $-e^{-1} < a < 0$. The recursion relation reads
\begin{equation}
\label{eq:second}
x_{1}=\exp(a), \quad x_{n+1} = \exp(a/x_{n}),~~n \geq 1.
\end{equation}
We employ again the contraction principle to prove that this recursion relation indeed converges to the right solution.

Let $X=[e^{-1},1]$ and $T \colon X \to X$ be the map $T(x)=\exp(a/x)$. We want to check whether
\begin{equation}
\label{eq:exey}
\abs{T(y)-T(x)} = \abs{e^{a/y}-e^{a/x}} \leq c\abs{y-x}
\end{equation}
with $c<1$. The function $c(a,x,y)=\abs{e^{a/y}-e^{a/x}}/\abs{y-x}$ is decreasing in all of its arguments, so its maximum should be at the point $Q=(e^{-1},e^{-1},-e^{-1})$. Unfortunately, $c(a,x,y)$ is undefined for $y=x$. However, since $c(a,x,y)=c(a,y,x)$, we argue that approaching $Q$ along the $x$-axis or the $y$-axis should give the same result, making $\lim_{P \to Q}c(P)$ well defined (if it exists at all). We then fix $x$ and let $y \to x$ to obtain
\begin{equation}
\lim_{y \to x}c(a,x,y) = \frac{\abs{ae^{a/x}}}{\abs{x^{2}}},
\end{equation}
where we resolved the indeterminate form $0/0$ by L'H\^{o}pital's rule. Setting $x=e^{-1}$ and $a=-e^{-1}$ we find that $c(a,x,y) \leq 1$, with strict inequality for $a>-e^{-1}$. We thus conclude that $T(x)$ satisfies the contraction principle on $X$ for $-e^{-1} < a < 0$ such that its unique fixed point on $X$ indeed solves Eq.~(\ref{eq:xlnxa}).

It is useful to examine an example. Let $a=-0.2$. Application of Eqs.~(\ref{eq:first}) provides the sequence $x_{1}=0.2 \to x_{2}=0.124267$ $\to \cdots \to x_{20}=0.0786584 \to x_{21}=0.0786584$, converging to seven decimal places in $20$ iterations, with $x_{20}\ln{x_{20}} \simeq -0.200\,000\,061$. The solution is $x_{l}^{*}=0.078\,658\,360\cdots$. Now let us apply Eqs.~(\ref{eq:second}) to find the righmost solution of Eq.~(\ref{eq:xlnxa}). We obtain $x_{1}=0.8187308 \to x_{2}=0.7832679 \to \cdots \to x_{11}=0.7716910 \to x_{12}=0.7716910$, and the procedure converges to seven decimal places in just $11$ iterations, with $x_{11}\ln{x_{11}} \simeq -0.199\,999\,981$. The precise value for the rightmost solution is $x_{r}^{*}=0.771\,690\,974\cdots$. The sequence converges faster because the Lipschitz constant for Eqs.~(\ref{eq:second}) is smaller than the one for Eqs.~(\ref{eq:first}), bringing separate points close together faster.

\subsubsection{\label{sec:other}Cases $a>0$}

When $a>0$, Sommerfeld splits the resolution of Eq.~(\ref{eq:xlnxa}) again into two cases. For $0<a<e$, the proper convergent iterative scheme reads the same as Eqs.~(\ref{eq:second}), namely,
\begin{equation}
\label{eq:0<a<e}
x_{1}=\exp(a), \quad x_{n+1} = \exp(a/x_{n}),~~n \geq 1,
\end{equation}
while for $a \geq e$ the scheme that converges coincides with Eqs.~(\ref{eq:first}) except that now we start with $x_{1}=a$, to wit,
\begin{equation}
\label{eq:a>e}
x_{1}=a, \quad x_{n+1} = \frac{a}{\ln{x_{n}}},~~n \geq 1.
\end{equation}
We invite the reader to repeat the analysis of the above maps along the lines of what we did for the other cases to prove their convergence, as well as to examine some examples. We will not make use of these procedures in our study.

In his mathematical memoir,\cite{Sommerfeld1898} Sommerfeld attempts a generalization of his analyses to an abstract equation of the form
\begin{equation}
\label{eq:general}
f(x)\phi(x)=a
\end{equation}
under mild hypotheses on the continuity of $f$, $\phi$, and their inverse functions. Barring possible factorizations or symmetrizations, a basic iterative scheme to solve Eq.~(\ref{eq:general}) either takes $f(x_{n+1})=a/\phi(x_{n})$ or $\phi(x_{n+1})=a/f(x_{n})$. Analysis of which scheme provides a contraction then suggests the procedure to follow.

\subsection{Solution of the approximate boundary condition}

We now apply what we have learned to the solution of Eq.~(\ref{eq:ulnuv}). We have a problem, though: equation Eq.~(\ref{eq:ulnuv}) involves complex quantities. The Banach contraction principle, however, only requires that the underlying space be complete in some norm, and this is the case of the complex plane---in fact, the complex plane $\mathbb{C}$ is isometric with the real plane $\mathbb{R}^{2}$, with the usual complex norm $\sqrt{z^{*}z}$ in the former corresponding naturally to the Euclidean norm $\sqrt{x^{2}+y^{2}}$ in the latter. Of course, before employing Eqs.~(\ref{eq:first}) or Eqs.~(\ref{eq:second}) one must analyse the moduli and the arguments of $x$ and $a$ to make a better informed trial.

Using the approximations $\mu \simeq \mu_{0}$ and $\eps\omega \ll \sigma$ valid for good metal conductors at low-frequencies (see the discussion preceding Eq.~(\ref{eq:kappa})) we obtain for $v$ (Eq.~(\ref{eq:uv}))
\begin{equation}
v \simeq \frac{e^{2\gamma}}{2i}\sqrt{\frac{\eps_{0}\omega}{i\sigma}}\,ka =
-\frac{e^{2\gamma}}{2\sqrt{2}}\sqrt{\frac{\eps_{0}\omega}{\sigma}}\,ka\,(1+i),
\end{equation}
and inserting the values of Sec.~\ref{sec:exact} for the 1~mm radius copper wire at $1$~GHz we get
\begin{equation}
\label{eq:v}
v \simeq -7.181 \times 10^{-7} (1+i),
\end{equation}
a number with a small negative real part. We shall thus try to apply procedure Eqs.~(\ref{eq:first}) to find the solution to Eq.~(\ref{eq:ulnuv}). We obtain, to three decimal places, the following sequence of numbers, where $z_{-e}$ reads $z \times 10^{-e}$: $u_{1}=(7.181+7.181\,i)_{-7}$ $\to u_{2}=(4.892+5.482\,i)_{-8}$ $\to \cdots$ $\to u_{6}=(4.095+4.529\,i)_{-8}$ $\to u_{7}=(4.095+4.529\,i)_{-8}$, and the procedure converged after just $6$ iterations to the solution
\begin{equation}
\label{eq:u}
u = (4.095+4.529\,i) \times 10^{-8}.
\end{equation}
This solution is slightly at variance with that found by Sommerfeld because he uses $v=-7.2\times 10^{-7}(1+i)$ and also because we keep more digits in the intermediate calculations.

From the value for $u$ above we obtain from Eq.~(\ref{eq:uv}) that
\begin{equation}
h = \sqrt{k^{2}+\frac{4e^{-2\gamma}}{a^{2}}u} = 20.960+1.362 \times 10^{-3}\,i.
\end{equation}
Compare this number with the one in Eq.~(\ref{eq:exacth}). The real parts are virtually identical (the difference starts from the fourth decimal place, not shown), but the imaginary parts differ almost by a factor of two. The root of this difference is likely the somewhat uncontrolled approximation $\sqrt{\hat{k}^{2}-h^{2}}/\hat{k} \approx 1$ that we made in passing from Eq.~(\ref{eq:trans}) to Eq.~(\ref{eq:rhs}); the approximation is valid in absolute value, but not componentwise. Following the analysis presented after Eq.~(\ref{eq:exacth}), we conclude that the wave is damped by a factor $e^{-1}$ every $\sim 734$~m and that
\begin{equation}
\frac{\omega}{h} = \frac{\omega}{k} \cdot \frac{k}{h} = c
[1-(5.878+6.500\,i) \times 10^{-5}],
\end{equation}
i.\,e., travels along the wire at a phase speed about $0.999\,941\,c$. While the physical conclusions in this case remain the same, namely, that the wave propagates reasonably undamped almost at the speed of light, the attenuation factor appears larger than it really is as a result of the approximations made.

In his textbook,\cite{Sommerfeld1952} Sommerfeld compares the above case with that of a very thin platinum wire of radius $a=2 \times 10^{-6}$~m forced at $f=300$~MHz. The conductivity of platinum at $293$~K is $\sigma=9.52 \times 10^{6}$~$\Omega^{-1}$\,m$^{-1}$, approximately $6.3$ times smaller than that of copper.\cite{Wikipedia2018} Note that now $\hat{\rho}(a)=\sqrt{\hat{k}^{2}-h^{2}}\,a$ is a small quantity and the approximation Eq.~(\ref{eq:rhs}) no longer holds---one has to expand the Bessel functions around $\hat{\rho}=0$ instead. The reader is invited to perform the analysis of the EM wave propagation characteristics in this case to conclude that the wire behaves almost like a purely resistive element, something that could be guessed already by a simple estimate of the skin depth $\delta(\omega) \sim \sqrt{2/\sigma\mu_{0}\omega}$.

%% %% %% %% %% %% %% %% %% %% %% %% %% %% %% %% %% %% %% %% %% %% %% %% %% %% %%

\section{\label{sec:lambert}The solution in terms of the \\ Lambert $W$ function }

The Lambert $W$ function has been around since at least the XVIII century, when it was introduced in disguise by Lambert in 1758 in the series solution of the trinomial equation $x=q+x^{m}$, a solution which Euler revisited and expanded in 1776. As applied mathematics and mathematical physics became more diverse in the XIX and XX centuries, the Lambert $W$ function started to pop up in several areas, from statistical mechanics and astrophysics to population dynamics, materials science, combinatorics, and the analysis of algorithms, among others.\cite{LambertW1993,LambertW1996,Valluri2000,Caillol2003,Cranmer2004,%
WarbWang2004,Packel2004,Hayes2005,Scott2006,Veberic2010} The fact that numerical routines to evaluate the Lambert $W$ function started to appear in the early 1970s also attests its ubiquitous occurrence in many areas of science and technology.\cite{Wright1959,Fritsch1973,Barry1995,Barry743,Boyd1998,Barry2000} By the mid-1990s, notation---in particular the use of the letter $W$ for the function after its first implementation in the Maple computer algebra software---and basic properties were settled, with Ref.~\onlinecite{LambertW1996} having become the standard reference on the Lambert $W$ function.

\begin{figure}[t]
\centering
\includegraphics[viewport=20 15 530 430, scale=0.40, clip]{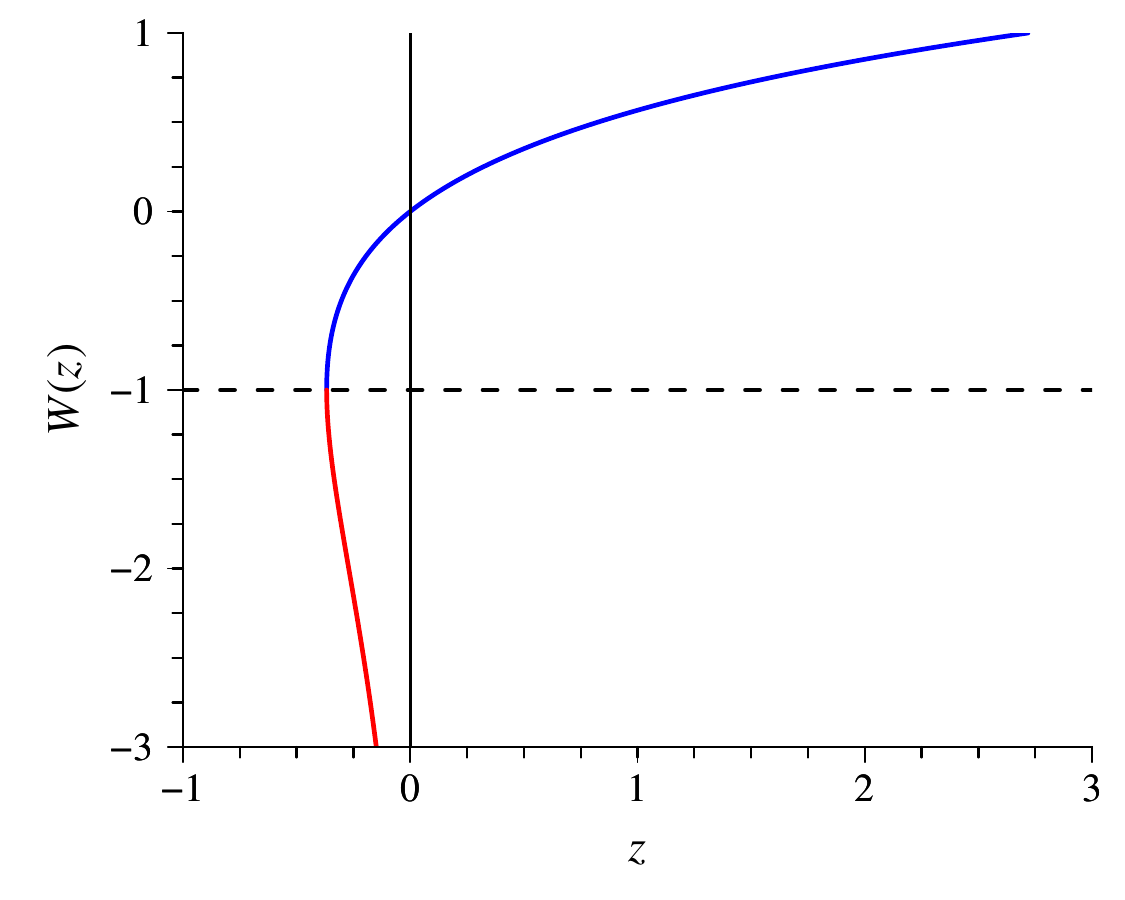}
\caption{The two branches of the Lambert $W$ function of a real argument $z \geq -e^{-1}$: $W_{0}(z) \geq -1$, the principal branch, and $W_{-1}(z) < -1$.}
\label{fig:LambW}
\end{figure}

The Lambert $W$ function is defined to be the inverse of the map $z \mapsto ze^{z}$, with $z$ a complex number. Put equivalently, the values of $W(z)$ are given by the solutions of the equation
\begin{equation}
\label{eq:lambW}
W(z)e^{W(z)} = z.
\end{equation}
Since the map $z \mapsto ze^{z}$ is not injective, the solutions of Eq.~(\ref{eq:lambW}) are multivalued. The map cannot be inverted easily---in fact, this is the whole point of defining the ``special function'' $W(z)$. A simple strategy to plot $W(x)$ for real $x$ is to plot $y=xe^{x}$ and interchange the roles of axis $x$ and $y$: look at the plot in a mirror or from behind and rotate it $\pi/2$ clockwise. A plot of $W(z)$ for real $z$ is displayed in Fig.~\ref{fig:LambW}. From this picture it is clear that the point $z=-e^{-1}$, at which $W(z)=-1$, is a branching point. The two real branches of $W(z)$ are the principal branch $W_{0}(z) \geq -1$, defined for all $z \geq -e^{-1}$, and $W_{-1}(z) < -1$, defined for $-e^{-1} \leq z < 0$ with a singularity $\lim_{z \to 0^{-}}W_{-1}(z) = -\infty$. If $z<-e^{-1}$ or $z$ is complex, then $W(z)$ becomes complex. The asymptotic behavior of $W_{0}(z)$ for small $z$ is given by
\begin{equation}
\label{eq:asympW}
W_{0}(z)=z-z^{2}+\frac{3}{2}z^{3}+\cdots, \quad \abs{z} \ll 1,
\end{equation}
while for large $z$ it is given by
\begin{equation}
\label{eq:bsympW}
W_{0}(z)=\ln{z}-\ln\ln{z}+\frac{\ln\ln{z}}{\ln{z}}+\cdots, \quad \abs{z} \gg 1.
\end{equation}

Back to our mathematical physics problem, if we put $z=\ln{u}$, Eq.~(\ref{eq:ulnuv}) becomes $ze^{z}=v$, which immediately furnishes the solution $u = e^{W(v)}$, or, equivalently,
\begin{equation}
\label{eq:elne}
u = \frac{v}{W(v)}.
\end{equation}
Unbeknownst to Sommerfeld, the solution of Eq.~(\ref{eq:ulnuv}) is ``simply'' given in terms of the Lambert $W$ function. This is not really helpful unless one has a procedure to compute $W(v)$---exactly what Sommerfeld devised with his iterative method. By comparing Figs.~\ref{fig:xlnxa} and \ref{fig:LambW} and noting that the two are connected by the relationship $z=\ln{x}$ one can easily conclude that the leftmost solution of Eq.~(\ref{eq:xlnxa}) corresponds to $\exp[W_{-1}(a)]$, while the righmost solution corresponds to $\exp[W_{0}(a)]$. In the software package Mathematica,\cite{Wolfram2018} one can calculate $W_{0}(z)$ with the command \verb#ProductLog[z]#, while the calculation of $W_{-1}(z)$ is achieved with the command \verb#ProductLog[1,z]#. When we insert $v = -7.181 \times 10^{-7}(1+i)$ (see Eq.~(\ref{eq:v})) into Eq.~(\ref{eq:elne}) we obtain, using Mathematica, the solution
\begin{equation}
\label{eq:umath}
u = (4.095+4.529\,i) \times 10^{-8},
\end{equation}
which is exactly the same value as before, Eq.~(\ref{eq:u}).

%% %% %% %% %% %% %% %% %% %% %% %% %% %% %% %% %% %% %% %% %% %% %% %% %% %% %%

\section{Summary and conclusions}

We have shown that a certain approximate equation arising from the boundary condition for EM waves between a cylindrical metallic wire and a dielectric can be solved by means of the Lambert $W$ function. Although one can solve the exact condition Eq.~(\ref{eq:trans}) numerically without much fuss, the physical rationale leading to the approximate condition Eq.~(\ref{eq:ulnuv}) is very illuminating and, as we have seen, provides a reasonably accurate description of the system. In fact, it is quite satisfying to verify that the answer obtained by Sommerfeld through his apparently offhand iterative method and the solution in terms of the Lambert $W$ function agree. While Sommerfeld's solution and the numerical solution of the exact boundary condition Eq.~(\ref{eq:trans}) do not quite match, they are qualitatively compatible and predict the same physics. We can trace the somewhat uncontrolled approximation $\sqrt{\hat{k}^{2}-h^{2}}/\hat{k} \approx 1$ as the probable reason for the mismatch between the exact and the approximate solutions.

Equation (\ref{eq:lambW}) is the simplest possible exponential polynomial equation, and this is arguably one of the reasons for its ubiquity. Slightly generalized equations, however, sometimes appear which can also be solved in terms of $W(z)$.\cite{Gosper1998} For example, in the study of the capture of a diffusing prey by diffusing predators in one dimension, the authors in
Ref.~\onlinecite{Sidney1999} obtain the equation $ye^{y^{2}}=M$, where $y=x_{\mathrm{last}}/\sqrt{4Dt}$ is the rescaled position of the last predator (the record position of the predators' one-dimensional random walks) and solve it iteratively as $y=\sqrt{\ln{M}}(1-\frac{1}{4}\ln\ln{M}/\ln{M}+\cdots)$. Clearly, squaring the original equation and then multiplying it by $2$ immediately leads to the solution $y^{2}=\frac{1}{2}W_{0}(2M^{2})$, from which good asymptotics, explicit numbers, and graphs can be obtained with a couple of commands in more or less any modern computer algebra system or scientific library.

The Lambert $W$ function can be computed to arbitrary precision by iterative root-finding of $we^{w}=z$, with the usual trade off between complexity of implementation and number of iterations to converge at the required precision. Newton's simple first-order method, for example, is appropriate, but converges too slowly for modern standards. The Maple software package implements Halley's third-order variant of Newton's method that attains high precision in affordable time.\cite{LambertW1993,Veberic2018} The investigation of the recursion relations (\ref{eq:first}) and (\ref{eq:second}) on the complex plane in the language of dynamical systems is not entirely without interest and may even prove useful in the implementation of numerical routines to calculate $W(z)$. An accessible and comprehensive account of discrete dynamical systems based on the ideas of inverse functions (``root-finding'') and fixed point theorems is given in Ref.~\onlinecite{Shlomo2013}.

In the introduction to Volume III of his Lectures on Theoretical Physics, Sommerfeld states that ``Heinrich Hertz's greatest paper on the `Fundamental Equations of Electrodynamics for Bodies at Rest' has served as model for my lectures on electrodynamics since my student days.''\cite{Sommerfeld1952} This is a remarkable note and advice, namely, that even the masters acknowledge that there is much to be learned from the masters. Maybe it is because they were closer to sparking gaps, vacuum tubes, flaring chemicals, and gears than the average contemporary student will ever be. There is a little bit---usually much more---for everyone. We encourage the reader to browse through Refs.~\onlinecite{Stratton1941,Sommerfeld1952,SommerfeldVI} looking for historical perspective, foundations, applications, and classical problems to solve.

%% %% %% %% %% %% %% %% %% %% %% %% %% %% %% %% %% %% %% %% %% %% %% %% %% %% %%

\begin{acknowledgments}

The author thanks the LPTMS for kind hospitality during a sabbatical leave in France and FAPESP (Brazil) for partial support through grant no. 2017/22166-9.

\end{acknowledgments}

%% %% %% %% %% %% %% %% %% %% %% %% %% %% %% %% %% %% %% %% %% %% %% %% %% %% %%

\end{document}